\documentclass[%
 reprint,
 twocolumn,
superscriptaddress,
showpacs,
 amsmath,amssymb,
 aps,
 prc,
]{revtex4-1}

\usepackage[pdftex]{graphicx}
\usepackage{dcolumn}
\usepackage{bm}
\usepackage{rotating}

\usepackage[T1]{fontenc}

\begin{document}

\preprint{APS/123-QED}

\title{Anomalous Neutron Yields Confirmed for Ba-Mo and Newly Observed for Ce-Zr from Spontaneous Fission of $^{252}$Cf}

\author{B. M. Musangu}
 \affiliation{Department of Physics and Astronomy, Vanderbilt University, Nashville, TN 37235, USA}
\author{A. H. Thibeault}
 \affiliation{Department of Physics and Astronomy, Vanderbilt University, Nashville, TN 37235, USA}

\author{T. H. Richards}
 \affiliation{Department of Physics and Astronomy, Vanderbilt University, Nashville, TN 37235, USA}
 \affiliation{Department of Physics and Astronomy, University of Alabama, Tuscaloosa, AL 35487, USA}

\author{E. H. Wang}
\author{J. H. Hamilton}
\author{C. J. Zachary}
\author{J. M. Eldridge}
\author{A. V. Ramayya}
 \affiliation{Department of Physics and Astronomy, Vanderbilt University, Nashville, TN 37235, USA}

\author{Y. X. Luo}
 \affiliation{Department of Physics and Astronomy, Vanderbilt University, Nashville, TN 37235, USA}
  \affiliation{Lawrence Berkeley National Laboratory, Berkeley, CA 94720, USA}
\author{J. O. Rasmussen}
 \affiliation{Lawrence Berkeley National Laboratory, Berkeley, CA 94720, USA}

\author{G. M. Ter-Akopian}
\author{Yu. Ts. Oganessian}
 \affiliation{Joint Institute for Nuclear Research, RU-141980 Dubna, Russian Federation}
 
 \author{S. J. Zhu}
 \affiliation{Department of Physics, Tsinghua University, Beijing 100084, China}

\date{\today}

\begin{abstract}
We reinvestigated the neutron multiplicity yields of Ba-Mo, Ce-Zr, Te-Pd, and Nd-Sr from the spontaneous fission of $^{252}$Cf; by (i) using both $\gamma$-$\gamma$-$\gamma$-$\gamma$ and $\gamma$-$\gamma$-$\gamma$ coincidence data, (ii) using up to date level scheme structures, and (iii) cross-checking analogous energy transitions in multiple isotopes, we have achieved higher precision than previous analyses. Particular attention was given to the Ba-Mo pairs where our results clearly confirm that the Ba-Mo yield data have a second hot fission mode where 8, 9, 10, and now 11 neutron evaporation channels are observed. These are the first observations of the 11 neutron channel. These 8-11 neutron channels are observed for the first time in the Ce-Zr pairs, but are not observed in other fission pairs. The measured intensities of the second mode in Ba-Mo and Ce-Zr pairs are $\sim$1.5(4)$\%$ and $\sim$1.0(3)$\%$, respectively. These high neutron number evaporation modes can be an indication of hyperdeformation and/or octupole deformation in $^{143-145}$Ba and in $^{146,148}$Ce at scission to give rise to such high neutron multiplicities.  
\end{abstract}

\maketitle


\section{\label{sec:level1}introduction}

Yields of individual correlated pairs in barium (Z = 56) and molybdenum (Z = 42) binary fission have been observed to undergo fission splits via an extra ``hot fission mode'' (also called second mode) \cite{Ter}. In this mode, it has been observed that the Ba-Mo fragment pair emits high neutron multiplicities of 7 to 10 neutrons in spontaneous fission of $^{252}$Cf \cite{Ter, Ako, Ta}. To explain this phenomenon, theorists have attributed the presence of this mode to a possible hyperdeformation of $^{144,145,146}$Ba fragments at scission \cite{Ter, Yu, Don}. This is justified by referring back to the theory which predicts that a large nuclear deformation is more likely to yield higher neutron multiplicities \cite{Bro}. Other theorists have raised skepticism, since the hot fission mode has only been observed in Ba-Mo fragment pairs of $^{252}$Cf and not in spontaneous fission of $^{248}$Cm \cite{Sch}. However, this private communication \cite{Sch} has never been published.

Furthermore, some earlier analysis in spontaneous fission of $^{252}$Cf did not confirm the second hot mode \cite{Bis} without reporting the 9 and 10 channel yields (see later discussion), while others did show some irregularity around the eight-neutron channel \cite{Wu, Swu, Goo, Ram}. Because of the importance of understanding this extra hot fission mode, pairs of Ba-Mo, Ce-Zr, Te-Pd, and Nd-Sr have been studied with improved precision using $\gamma$-$\gamma$-$\gamma$-$\gamma$ as well as $\gamma$-$\gamma$-$\gamma$ coincidence data and the latest level structures of these nuclei. Also, relative intensities of transitions in these nuclei made available through our work likewise improved the accuracy of the analysis. In all cases, careful attention was given to transitions of the same energies in multiple isotopes.

Of particular interest in this experiment are the $\gamma$-ray transitions to the ground state. Some isotopes have a single ground state $\gamma$-ray transition, but others have multiple ones. The ground state $\gamma$-ray transition is generally the highest intensity $\gamma$-ray emitted by an isotope, and all daughter nuclei will emit this $\gamma$-ray, excluding the extremely unlikely case they were produced in the ground state during the fission process. By measuring the intensity of ground state $\gamma$-rays, it can be deduced how likely specific isotopes of fission partner isotopes are to be produced in the spontaneous fission of $^{252}$Cf.

If the ground state $\gamma$-ray is inconvenient to measure, it is also possible to use a higher transition to make this calculation, as long as its intensity relative to the ground state transition is known. The relative intensities of all transitions feeding the ground states of the isotopes analyzed in this study were determined based on new levels schemes with new ground state transitions (especially in odd-even nuclei) \cite{Urb, Sim, Mar, Urb1, Urb2, Urb3, Urb4, Din, Luo} to produce a new set of absolute yields. Some of the $\gamma$-rays in the newly published level schemes are not clearly observed in our data such as $^{140}$Te \cite{Mo}. The new results confirm a second hot mode in Ba-Mo pairs with an intensity of $\sim$1.5(4)$\%$ and shows evidence for a comparable second hot mode in Ce-Zr pairs with an intensity of $\sim$1.0(3)$\%$. These result are compared with other results \cite{Bis, Wu, Swu, Goo, Ram}.

\section{experimental set-up}

The present experiment was done at the Lawrence Berkeley National Laboratory (LBNL) with the Gammasphere detector array. A 62 $\mu$Ci $^{252}$Cf source was sandwiched between two iron 10 mg/cm$^{2}$ foils, which were used to stop the fission fragments and eliminate the need for Doppler correction. A 7.62 cm in diameter plastic (CH) ball surrounded the source to absorb $\beta$ rays and conversion electrons, as well as to partially moderate and absorb fission neutrons. A total of 5.7$\times$10$^{11}$ $\gamma$-$\gamma$-$\gamma$ and higher fold $\gamma$ events, and 1.9$\times$10$^{11}$ $\gamma$-$\gamma$-$\gamma$-$\gamma$ and higher fold $\gamma$ coincident events were recorded. These $\gamma$ coincident data were analyzed by the RADWARE software package \cite{Rad}. More details about the experimental setup can be found in Refs.~\cite{Ham,Wan15}.

\section{method of data analysis}

Quadruple ($\gamma$-$\gamma$-$\gamma$-$\gamma$) as well as triple ($\gamma$-$\gamma$-$\gamma$) coincidence data were analyzed to extract the relative yields of correlated fragment pairs in spontaneous fission of $^{252}$Cf. In order to find peaks for the yield computation, a double or triple gate was set on the most intense coincident $\gamma$-rays in a given nucleus (usually the 2$^{+}$ $\rightarrow$ 0$^{+}$ and 4$^{+}$ $\rightarrow$ 2$^{+}$ transitions in case of an even-even product). On the generated coincidence spectrum, the transitions in the partner fragments were clearly identified. The intensities of the $\gamma$-ray transitions in the partners (usually the 2$^{+}$ $\rightarrow$ 0$^{+}$ in case of even-even nuclei) were corrected for the detector efficiencies and internal conversion coefficients (ICC) of the $\gamma$-rays involved in the selection and used along with other transitions feeding into the ground state to extract the relative yields for the considered partitions. In the case of odd nuclei, all the known transitions populating the ground state were summed proportionally according to their intensities.

Additionally, if there is a presence of an isomeric state in the level scheme structure of a given nucleus, the transitions populating into that isomeric state were considered by adding the contribution of those transitions populating that state according to its time scale. This was done to avoid underestimating the yields. Specific examples will be given in the  discussion section. A two-dimensional matrix was created from the initial data by selecting the $\gamma$-ray coincidences occurring within 1 $\mu$s time window. The peaks observed in this two-dimensional spectrum arise from the coincidences between the $\gamma$-ray emitted promptly by both complementary fission fragments of different fragment pairs.

\section{Experimental results and discussion}

Fission spectra are very complex and this type of analysis is difficult and prone to errors caused by random coincidences and background. As such, we found some peaks unusable because of contamination or similar transition energies found in other isotopes. Cross-checks by gating on a series of isotopes as well as gating on their fission partners have been done to determine possible contamination and the accuracy of the current result. In addition, to measure yields in these cases, we used peaks found in higher transitions and scaled them appropriately. For example, Table~\ref{table-of-scaling-factors} contains this information for the Sr-Nd pair. In order to calculate the scaling factor, we set a clean gate with no contamination on energy transitions of other isotopes and measured the intensities of the ground state transitions and of the higher transitions. By taking the ratio of these intensities, we compute the scaling factor needed.

\begin{table}[htp!]
\centering\caption{\label{table-of-scaling-factors}A list of isotopes whose ground state transition energies were difficult to measure (because of similar ground state energies or not clearly observed in our data) and what energy transition we measured instead in Nd-Sr fragment pairs. The scaling factor is the relative intensity of the measured transition to the ground state transition; we divided the yield of the transition by this factor to correct it. }
\resizebox{\columnwidth}{!}{
\begin{tabular}{|c|l|c|c|c|}
\hline
{Isotope}                     & Gate           & Ground   & Measured    & Scaling  \\ 
                              &                & state (keV) & (keV)          & factor      \\ \hline
\rule{0pt}{2.5ex}$^{92}$Sr    & $^{155}$Nd & 814.6       & 859            & 0.46           \\ \hline
\rule{0pt}{2.5ex}$^{94}$Sr    & $^{150}$Nd     & 836.7       & 1089.1         & 0.23           \\ \hline
\rule{0pt}{2.5ex}$^{96}$Sr    & $^{155}$Nd     & 814.8       & 977.5          & 0.49           \\ \hline
\end{tabular}}
\end{table}

\begin{figure}[htp!]
    \centering
    \includegraphics[width=\columnwidth]{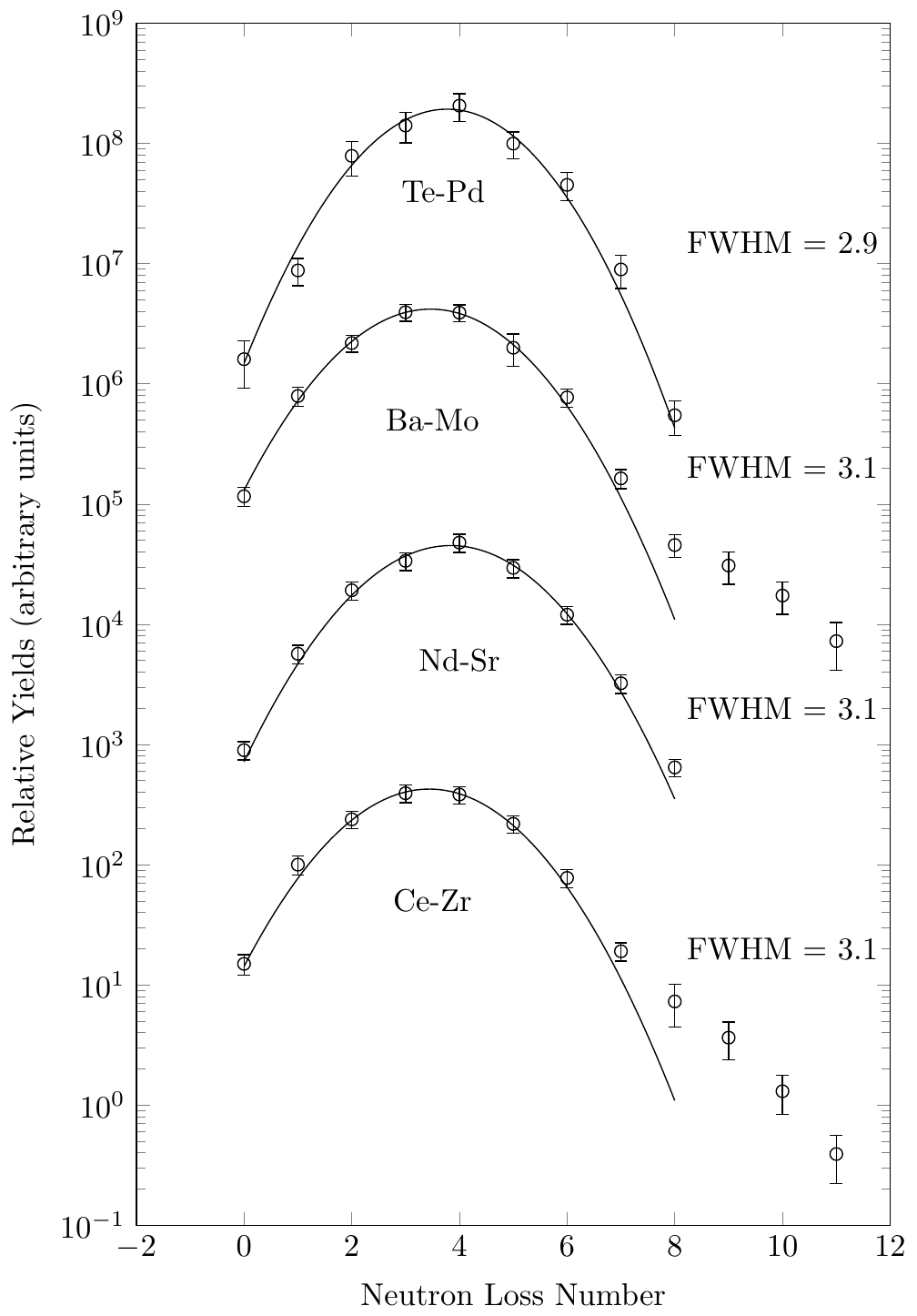}
    \caption{\label{yield-curves}The experimental Ba-Mo, Ce-Zr, Te-Pd and Nd-Sr yield curves from the present analysis are shown above. There is no evidence for the 9, 10 and 11 neutron channels in pairs other than Ba-Mo and the newly observed Ce-Zr. A smooth Gaussian fit to the 0-8 neutron channels in Nd-Sr, Te-Pd, Ce-Zr and Ba-Mo. The full width at half maximum (FWHM) was also calculated for each of the yields as shown in this figure.}
    \label{fig:my_label}
\end{figure}

\begin{figure}[ht!]
    \centering
    \includegraphics[width=\columnwidth]{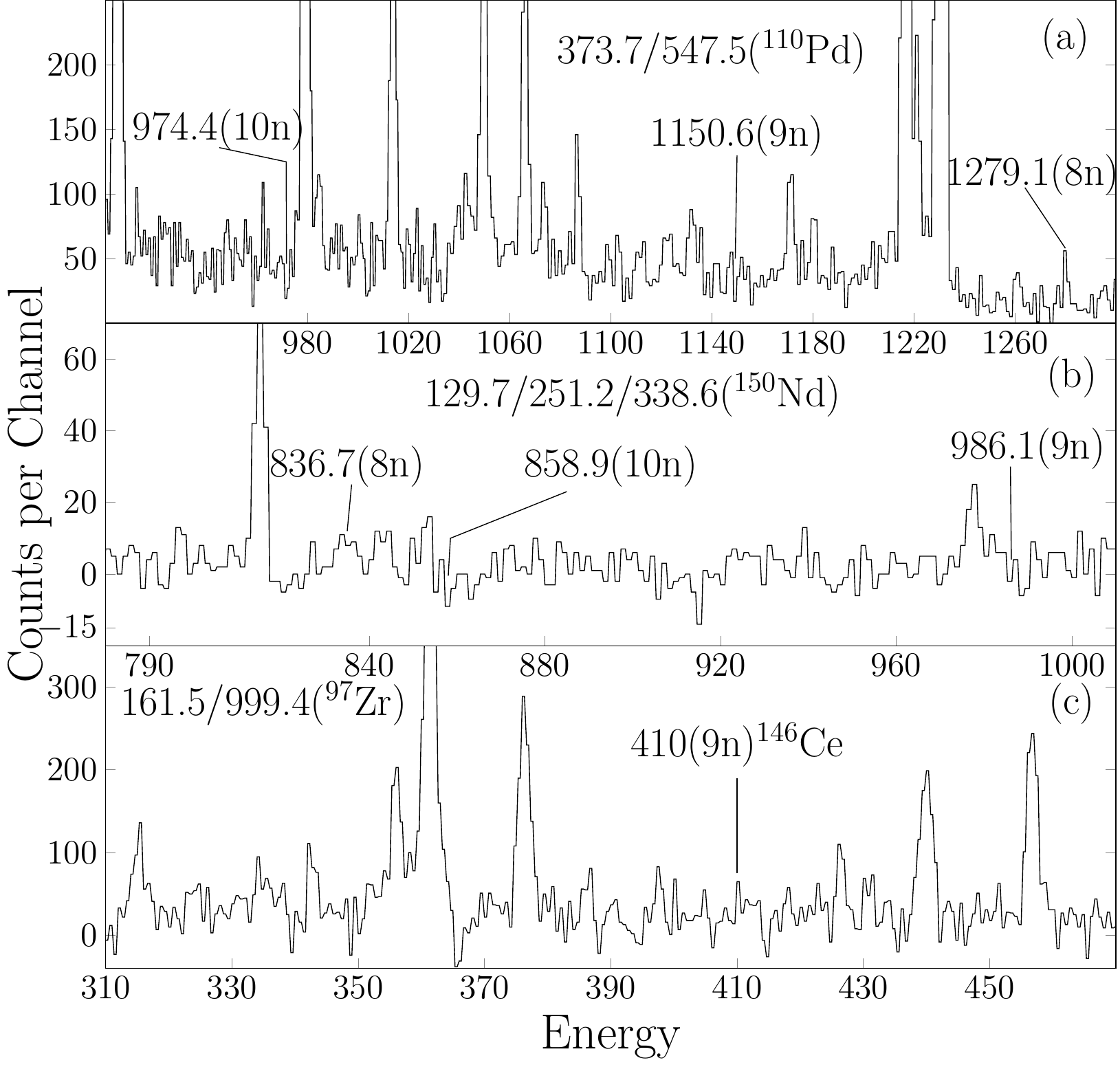}
    \caption{\label{no-evidence} Gamma-ray coincidence spectra by gating on (a) 373.7 and 574.5 keV transitions in $^{110}$Pd to show that there is no evidence for the 9 and 10 neutron channel at 1150.6 keV in $^{133}$Te and at 974.4 keV in $^{132}$Te, respectively, whereas there is clear evidence of the 8 neutron channel at 1279.1 keV in $^{134}$Te which fits to the curve. In (b) a triple gate on 129.7, 251.2 and 338.6 keV transitions in $^{150}$Nd to show that there is no evidence for the 9 and 10 neutron channel at 986.1 keV in $^{93}$Sr and at 858.9 keV in $^{93}$Sr for the Nd-Sr pair, respectively, whereas there is weak evidence of the 8 neutron channel at 836.7 keV in $^{94}$Sr. And in (c) a double gate on 161.5 and 999.4 keV transitions in $^{97}$Zr to show evidence for the 9 neutron channel at 410 keV in $^{146}$Ce for the Ce-Zr pair.}
    \label{fig:my_label1}
\end{figure}

The yield matrix for tellurium (Z = 52) and palladium (Z = 46) in Table~\ref{Te-Pd-yield-table}, displays the expected pattern where the highest yields are concentrated in the center of the matrix, along the 4 neutron channel diagonal, running from the bottom left to the top right corners. The yield matrix was normalized by using the normalization constant of $^{114}$Pd in Wahl's table \cite{Wahl}. This pair has a lot of isomeric states and many of the level schemes are incomplete. Therefore, the yields are incomplete (see Fig.~\ref{yield-curves}). The yields for the other studied element pairs in Fig.~\ref{yield-curves} display a similar pattern as expected. However, as seen in Fig.~\ref{no-evidence} (a), where the spectrum was gated on both the 373.7 and 574.5 keV transitions in $^{110}$Pd, there is no evidence for the 9 and 10 neutron channel at 1150.6 keV in $^{133}$Te and at 974.4 keV in $^{132}$Te, respectively, for the Te-Pd pairs. Whereas, there is clear evidence of the 8 neutron channel at 1279.1 keV in $^{134}$Te which fits nicely with the simple curve in Fig.~\ref{yield-curves}. 

\begin{table*}[htp!]
\caption{\label{Te-Pd-yield-table}New yield matrix for tellurium and palladium from the spontaneous fission of $^{252}$Cf.}
\begin{ruledtabular}
\begin{tabular}{l|lllllllll}
 {Yield}     & \,$^{110}$Pd     & $^{111}$Pd     & $^{112}$Pd    & $^{113}$Pd   & $^{114}$Pd    & $^{115}$Pd   & $^{116}$Pd    & $^{117}$Pd    & $^{118}$Pd    \\ \hline
\rule{0pt}{2.5ex}$^{130}$Te &           &           &          &         &          &         & 0.010(1) & 0.0029(4) & 0.017(2) \\
$^{131}$Te &           &           &          &         &          & 0.05(1) & 0.042(5) & 0.006(1) & 0.010(1) \\
$^{132}$Te &           &           &          & 0.036(4) & 0.06(1)  & 0.07(1) & 0.19(2)  & 0.018(2) & 0.016(2) \\
$^{133}$Te &           & 0.0022(3) & 0.016(3) & 0.035(4) & 0.14(2)  & 0.09(1) & 0.15(2)  & 0.005(1) & 0.008(1) \\
$^{134}$Te & 0.0021(3) & 0.015(2)  & 0.16(2)  & 0.33(4) & 0.78(9) & 0.28(3) & 0.24(3)  & 0.007(1) & 0.006(1) \\
$^{135}$Te & 0.003(1)  & 0.017(2)  & 0.11(1)  & 0.20(2) & 0.27(4)  & 0.05(1) & 0.040(5) & 0.0013(3) &          \\
$^{136}$Te & 0.021(3)  & 0.06(1)   & 0.29(4)  & 0.24(3) & 0.22(4)  &         &          &          &          \\
$^{137}$Te & 0.012(2)  & 0.020(3)  & 0.12(2)  & 0.06(1) & 0.013(2) &         &          &          &          \\
$^{138}$Te & 0.014(2)  & 0.023(3)  & 0.035(4)  &         & 0.005(1) &         &          &          &          
\end{tabular}
\end{ruledtabular}
\end{table*}

\begin{table}[htp!]
\centering\caption{\label{table-of-ave-neu}A list of the average neutron multiplicities (${\bar{\nu}}$) and the full width at half maximum (FWHM) for each pair shown in Fig.~\ref{yield-curves}. The average neutron multiplicity distributions are very close to the accepted values of 3.8 for the spontaneous fission of $^{252}$Cf. }{
\begin{tabular}{|c|l|c|c|c|}
\hline
                                       & Ba-Mo           & Ce-Zr      & Te-Pd       & Nd-Sr  \\ \hline
\rule{0pt}{2.0ex}Ave (${\bar{\nu}}$)   & 3.57           & 3.62        & 3.71           & 3.84 \\ \hline
\rule{0pt}{2.0ex}FWHM                  & 3.1            & 3.1         & 2.9            & 3.1           \\ \hline
\end{tabular}}
\end{table}

In the Nd-Sr yields matrix in Table~\ref{nd-sr}, we left most of the items in the $^{148}$Nd and $^{149}$Nd columns blank because those isotopes are very weakly populated in the spontaneous fission of $^{252}$Cf, making it difficult to measure energy transitions of interest. Some similar energies, such as the ground state transitions of $^{100}$Sr and $^{150}$Nd (129.8 keV and 129.7 keV, respectively), are not distinguishable. To resolve this, we set a gate on both the 2$^{+}$ $\rightarrow$ 0$^{+}$ and 4$^{+}$ $\rightarrow$ 2$^{+}$ transitions of $^{100}$Sr 129.8/287.9 keV and compared the intensity ratios of its partners' transitions of interest to another double gate on 2$^{+}$ $\rightarrow$ 0$^{+}$ and 4$^{+}$ $\rightarrow$ 2$^{+}$ transitions of $^{98}$Sr 144.3/289.4 keV. The 129.7 keV peak from $^{150}$Nd was contaminated as well as some of the other peaks of interest. Therefore, the gate on $^{100}$Sr was avoided. Isomeric states had to be considered in the case of $^{153,154}$Nd. In $^{154}$Nd, the isomeric state at 1298.0 keV is weakly populated therefore, the transitions feeding into it were not added to the total yields.

The $^{154}$Nd nucleus was reported to have an isomer at 1348 keV in \cite{Gau}, which is not observed in the current data. In that paper, the ground state band transitions were reported as 72, 163, 243, 328 keV... etc, with a 870 keV isomeric transition. In the contrast, both the previous work in \cite{Urb5} and our current data show a 72-163-248 keV cascade for ground state band. In \cite{Gau}, transition energy and levels were also reported in other nuclei $^{156,158}$Sm, $^{152,156}$Nd. The energy difference between transitions in those nuclei reported in \cite{Gau} and our current work, as well as other data recorded in nuclear data sheets is generally within 1 keV. Thus, the big ~5 keV energy difference in $^{154}$Nd between the 243 and 248 keV 6$^{+}$ $\rightarrow$ 4$^{+}$ transition may indicate a wrong isotope assignment in \cite{Gau}. Instead, $^{159}$Sm was reported to have an isomer in \cite{Sim1}, with 163-243 keV for the first two E2 transitions for the ground state band and 870 keV for the isomeric transition. The 1348 keV isomer reported in \cite{Gau} may belong to $^{159}$Sm, but 5 mass number away from $^{154}$Nd. Further details are needed to understand the reason.

According to \cite{Sim}, $^{153}$Nd the ground state transitions are 50.0 keV, 120.2 keV and 191.7 keV if our time gate is long enough to cover the isomeric transition. Energies at 50.0 keV, 70.2 keV, 60.7 keV and 78.0 keV reported in \cite{Sim} are hard to measure accurately to get accurate intensities. Thus, when the ground state transitions are hard to measure we summed up all the next level transitions. In the case of $^{153}$Nd, we used 88.3 keV, 197.6 keV and 158.5 keV in the ground state band and 97.9 keV, 175.8 keV and 208.8 keV in the 5/2$^+$ band together. These transition are reported in \cite{Sim}. Figure.~\ref{yield-curves} shows a plot of the extracted yields against the fission's neutron channel number (see Fig.~\ref{yield-curves}). Also shown in Fig.~\ref{no-evidence} (b), a triple gate on 129.7, 251.2 and 338.6 keV transitions in $^{150}$Nd is used to show that there is no evidence for the 9 and 10 neutron channel at 986.1 keV in $^{93}$Sr and at 858.9 keV in $^{93}$Sr for the Nd-Sr pair, respectively. Whereas there is clear evidence of the 8 neutron channel at 836.7 keV in $^{94}$Sr that fits fits nicely the single yields curve as shown in Fig.~\ref{yield-curves}.

\begin{table*}[tp!]
\caption{\label{nd-sr}New yield matrix for neodymium and strontium from the spontaneous fission of $^{252}$Cf.}
\begin{ruledtabular}
\begin{tabular}{l|lllllllll}
{Yield} & $^{148}$Nd    & $^{149}$Nd    & $^{150}$Nd    & $^{151}$Nd    & $^{152}$Nd    & $^{153}$Nd    & $^{154}$Nd    & $^{155}$Nd    & $^{156}$Nd    \\ \hline
\rule{0pt}{2.5ex}$^{91}$Sr      &          &          &          &          &          &          & 0.004(1) &          & 0.005(1) \\
$^{92}$Sr      &          &          &          &          &          &          & 0.020(3) & 0.019(4) & 0.030(5) \\
$^{93}$Sr      &          &          &          &          & 0.005(1) & \textless{}0.05         & 0.027(5) & 0.012(2) & 0.022(4) \\
$^{94}$Sr      &          &          & 0.010(2) & 0.026(4) & 0.09(1)  & 0.18(3)  & 0.13(2)  & 0.047(8) & 0.043(7) \\
$^{95}$Sr     &          &          & 0.017(3) & 0.033(6) & 0.12(2)  & 0.30(5)  & 0.11(2)  & 0.023(4) & 0.012(2) \\
$^{96}$Sr     &          &          & 0.040(7) & 0.06(1)  & 0.16(3)  & 0.23(4)  & 0.06(1)  & 0.010(2) & 0.005(1) \\
$^{97}$Sr     &          &          & 0.030(5) & 0.043(7) & 0.06(1)  & 0.07(1)  & 0.021(4) &          &          \\
$^{98}$Sr      & 0.010(2) & 0.031(6) & 0.08(2)  & 0.044(7) & 0.06(1)  & 0.025(5) & 0.009(2) &          &          \\
$^{99}$Sr      &          &          & 0.020(4) & 0.017(3) & 0.015(3) &          &          &          &          \\ 
$^{100}$Sr     &          &          & 0.026(4) & 0.006(1) &          &          &          &          &          
\end{tabular}
\end{ruledtabular}
\end{table*}

\begin{table*}[t!]
\caption{\label{ce-zr}New yield matrix for cerium and zirconium from the spontaneous fission of $^{252}$Cf. The 8-11 neutron channels are labeled with neutron numbers as superscripts.}
\begin{ruledtabular}
\begin{tabular}{l|lllllllll}
{Yield}                         & $^{144}$Ce & $^{145}$Ce& $^{146}$Ce& $^{147}$Ce& $^{148}$Ce& $^{149}$Ce& $^{150}$Ce& $^{151}$Ce & $^{152}$Ce    \\ \hline
\rule{0pt}{2.5ex}$^{96}$Zr      &                 &          & 0.003(1)$^{10}$ & 0.004$^{9}$          & 0.004(1)$^{8}$ &          & 0.015(3) & 0.025(5) & 0.020(4) \\
$^{97}$Zr                       & 0.002(1)$^{11}$ & 0.005(2)$^{10}$& 0.003(1)$^{9}$ & \textless{}0.009$^{8}$ & 0.028(4) & 0.026(5) & 0.07(1)  & 0.08(2)  & 0.050(9) \\
$^{98}$Zr                       & 0.004(1)$^{10}$ & 0.005(2)$^{9}$ & 0.008(3)$^{8}$& 0.048(8)& 0.06(1)  & 0.14(3)  & 0.21(4)  & 0.15(3)  & 0.06(1)  \\
$^{99}$Zr                       & 0.007(3)$^{9}$ & 0.006(2)$^{8}$ & 0.018(3) & 0.056(9)  & 0.20(3)  & 0.25(5)  & 0.30(5)  & 0.12(2)  & 0.033(6) \\
$^{100}$Zr                      & 0.011(2)$^{8}$ & 0.032(6) & 0.12(2)  & 0.28(4)   & 0.55(9)  & 0.39(7)  & 0.28(5)  & 0.11(2)  & 0.017(3) \\
$^{101}$Zr                      &          & 0.11(2)  & 0.26(4)  & 0.33(5)   & 0.54(9)  & 0.23(4)  & 0.17(3)  &          &          \\
$^{102}$Zr                      & 0.024(4) & 0.16(3)  & 0.43(7)  & 0.40(6)   & 0.40(6)  & 0.10(2)  & 0.029(5) &          &          \\
$^{103}$Zr                      & 0.033(6) & 0.14(3)  & 0.22(4)  & 0.14(3)   & 0.12(2)  & 0.03(1)         &          &          &          \\
$^{104}$Zr                      & 0.005(1) & 0.07(1)  & 0.008(1) & 0.0013(3) &          &          &          &          &          
\end{tabular}
\end{ruledtabular}
\end{table*}

In the case of determining the cerium (Z = 58) and zirconium (Z = 40) yield matrix, measurements of multiple $\gamma$-rays emitted by the Ce-Zr fission fragment pairs formed in spontaneous fission of $^{252}$Cf were used to extract the yields. Table~\ref{ce-zr} below displays the absolute yields data that were collected. These are new results and different from the report given in Ref.~\cite{brok}. In this analysis, most of the transitions of interest were easily identifiable with the exception of the 97.5 keV (9/2$^-$ $\rightarrow$ 5/2$^-$) and 97.4 keV (2$^{+}$ $\rightarrow$ 0$^{+}$) from $^{145}$Ce and $^{150}$Ce, respectively, and $^{101}$Zr and $^{103}$Zr also have similar transitions of 97.8 keV (5/2$^{+}$ $\rightarrow$ 3/2$^{+}$) and 98.4 keV (5/2$^{+}$ $\rightarrow$ 3/2$^{+}$), respectively.

 \begin{figure}[!ht]
    \centering
    \includegraphics[width=\columnwidth]{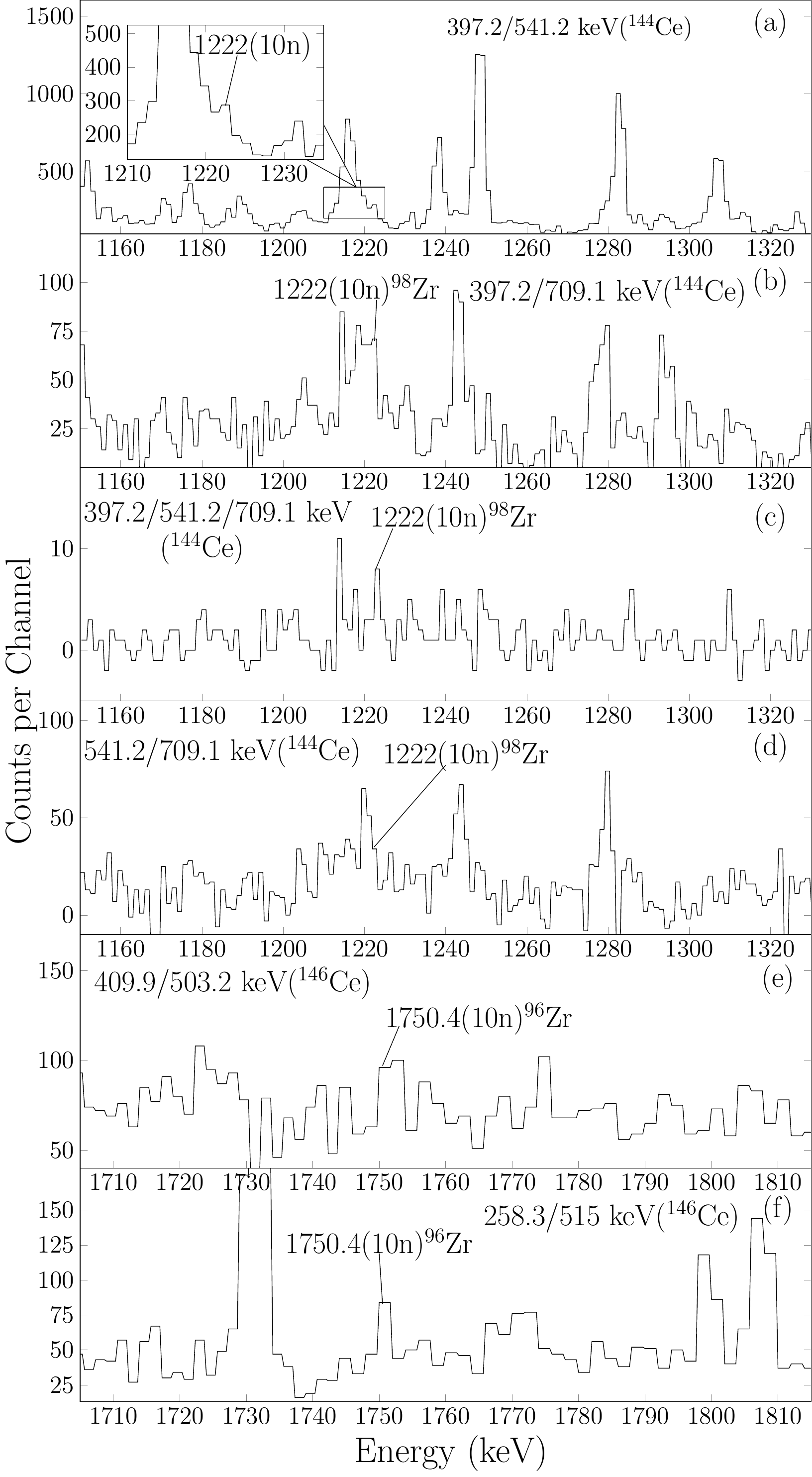}
    \caption{\label{evidence_for_ce} Gamma-ray coincidence spectra by gating on (a) 397.2/541.2 keV, (b) 397.2/709.1 keV, (c) 397.2, 541.2 and 709.1 keV and (d) 541.2/709.1 keV transitions in $^{144}$Ce to show that there is evidence for the 10 neutron channel at 1222 keV in $^{98}$Zr for the $^{98}$Zr-$^{144}$Ce pair. In (e), a double gate on 409.9/503.2 keV transitions in $^{146}$Ce to show clear evidence of the 10 neutron for the $^{146}$Ce and $^{96}$Zr pair at 1750.4 keV by gating on 409.9/503.2 keV. And (f) gives further evidence for the presence of the 1750.4 keV transition by gating on 258.3/515 keV}
    \label{fig:my_label1}
\end{figure}

To avoid possible contamination, a few gates were set on $^{150}$Ce to measure the peaks of interests from its Zr fragment partners. Any contamination of the 98.4 keV transition from $^{103}$Zr is avoided since $^{103}$Zr and $^{150}$Ce are not fission partners in spontaneous fission of $^{252}$Cf. However, gating on $^{145}$Ce would bring in contamination from both $^{101}$Zr and $^{103}$Zr. Multiple gates were set on the Zr fragments to measure the ground state transition of $^{145}$Ce. By gating on 109.4/146.6 keV of $^{103}$Zr, the 97.4 keV transition from $^{150}$Ce is once more avoided. Because the channel number between $^{102}$Zr and $^{150}$Ce is zero, any possible contribution from the 97.4 keV of $^{150}$Ce to the 97.5 keV in $^{145}$Ce can be neglected given that it is very small. Any gate on $^{101}$Zr brings in contribution from both the 97.4 keV of $^{150}$Ce to the 97.5 keV in $^{145}$Ce. One thing to consider first is to avoid setting any gate using the 97.8 keV in $^{101}$Zr. This prevents the contribution from the 98.4 keV in $^{103}$Zr. A double gate 216.6/250.9 keV was set on $^{101}$Zr and from this gate three peaks were of interest, however, only the 158.7 keV one in $^{150}$Ce and a peak around 98 keV, containing both the 97.4 keV peak ($^{145}$Ce) and 97.5 keV peak ($^{150}$Ce), were capable of being measured. Since the ratio between 158.7 keV peak ($^{145}$Ce) and 97.5 keV peak ($^{150}$Ce) was already known when determining the yields of $^{150}$Ce, it was easy to deduce the portion of the contribution of $^{150}$Ce from the measured peak. This meant that the remaining portion belonged to the 97.4 keV peak ($^{145}$Ce). This method was repeated for $^{100}$Zr.

In the present study we observed evidence of the 9, 10 and 11 neutron channels in the Ce-Zr fission pairs. A double gate on 397.2 keV and 541.2 keV in $^{144}$Ce shows evidence for the 10 neutron channel at 1222.9 keV in $^{98}$Zr (see Fig.~\ref{evidence_for_ce}). However, in this gate the intensity is not very clear. Therefore, we checked for clean transitions to gate on in $^{144}$Ce to avoid contamination and we found that other double and triple gates on 397.2, 541.2, 709.1 and 585.2 keV were good candidates that can be used to verify this observed peak. As shown in Fig.~\ref{evidence_for_ce} (a), (b), (c) and (d) all these gates show evidence of the presence of the 10 neutron channel for the $^{98}$Zr-$^{144}$Ce fission pair. In part (e) of Fig.~\ref{evidence_for_ce} there is another clear evidence of the 10 neutron for the $^{146}$Ce and $^{96}$Zr pair at 1750.4 keV by gating on 409.9/503.2 keV in $^{146}$Ce. To measure the intensity of the 1750.4 keV, we accounted for the presence of 1751 keV in $^{100}$Zr (which would make this yield higher than what it should be) by subtracting the portion of 1751 keV from the measured 1750.4 keV in $^{146}$Ce and $^{96}$Zr pair since we already had the real intensity for that. Part (f) of Fig.~\ref{evidence_for_ce} gives further evidence for the presence of the 1750.4 keV.

There is a clear 1103 keV peak when gating on $^{146}$Ce. The intensity of this peak however, is higher than expected. We discovered that this high intensity is due to a strong contamination around this peak from beta decay where a 1103 keV transition in $^{146}$Ce feeds a 1810.2 keV level. There is also a clear peak at 258 keV in $^{146}$Ce when one gates on $^{97}$Zr. However, for any of the possible gates on $^{97}$Zr, it is difficult to find a good reference peak that has the expected ratio with the 258 keV peak. Therefore, we measured the 409.9 keV peak which has the expected ratio with the 209.1 keV peak taking into consideration their intensities, efficiency, and internal conversion relative to the ground state transitions. Another challenging channel to measure is the $^{144}$Ce-$^{97}$Zr. There is a strong 1102.8 keV peak feeding into the 4$^{+}$ level (938.6 keV) in $^{144}$Ce. Hence, the presence of the 11 neutron channel at about the 397.2 keV peak would be influenced by the overlapping two transitions of 1103 keV in both $^{144}$Ce and $^{97}$Zr.

\begin{figure}[ht!]
    \centering
    \includegraphics[width=\columnwidth]{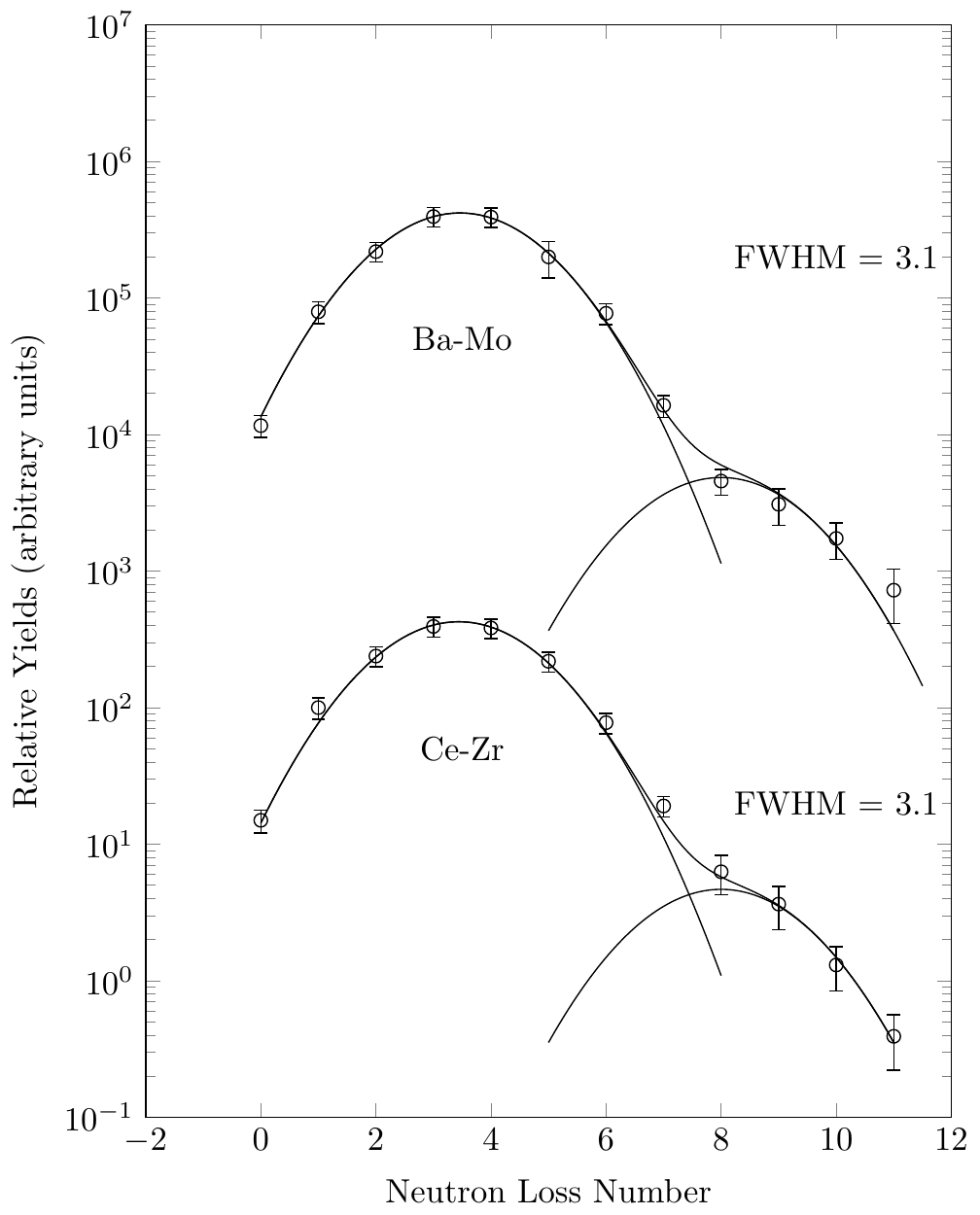}
    \caption{\label{tailscurves}The second curve in Ce-Zr was fitted by fixing the width of the second curve (presenting the second mode) to the width of the first curve (presenting the first modes) and also fixing the position to 8 neutron channel. It contributes $\sim$1.0(3)$\%$ of the first mode. The second curve in the Ba-Mo fit was also fitted by fixing the width of the second curve to the width of the first curve and fixing the position to 8 neutron channel. It contributes $\sim$1.5(4)$\%$ of the first mode.}
    \label{fig:my_label1}
\end{figure}

Upon completion of the matrix yield of the correlated fragment pairs of Ce-Zr in the spontaneous fission of $^{252}$Cf, the yields were next scaled according to Ter-Apkopian's independent yield~\cite{Ako} and summed for each isotope of Ce. This summation and Ter-Akopian's calculated data for Ce-Zr, were both normalized such that $^{148}$Ce had a value of 100. Then these two data sets were compared to see if Ter-Akopian's calculations could be verified. As can be seen in Fig.~\ref{yield-curves}, the present absolute yield data Te-Pd, Nd-Sr and Ce-Zr are in agreement with the previous ones~\cite{Ako}, and thus are experimentally confirmed with smaller error limits. The results from the present study show evidence for an ``extra hot fission mode" as shown Fig.~\ref{tailscurves}. This is the first time this mode is observed in Ce-Zr pairs; it is $\sim$1.0(3)$\%$ of the first mode. The observation of this mode in this pair can be explained when one considers that $^{143-145}$Ba and $^{146,148}$Ce have been determined to be octupole deformed \cite{Ham1, Buc, Enh, Chen, Zhu} and may also have hyperdeformation at scission to give these nuclei high internal energy and in turn gives rise to high neutron multiplicities. The second curve (6-10 neutrons) in Fig~\ref{tailscurves} was fitted by restricting the width of the second curve to the width of the first curve (0-7 neutron channels) and the position to 8 neutron channel. If the unfixed width method is used instead, the width of the Ce-Zr first curve is 6$\%$ larger than the Ba-Mo width. However, the 10 neutron channel in Ce-Zr pair is obviously above the tail of the first Gaussian in either way.

  \begin{table*}[!htp]
 \caption{\label{mp-ba}New yield matrix for barium and molybdenum from the spontaneous fission of $^{252}$Cf. The 8-11 neutron channels are labeled with neutron numbers as superscripts.}
 \begin{ruledtabular}
\begin{tabular}{l|lllllllllll}
{Yield}                         & 138Ba            & $^{139}$Ba     & $^{140}$Ba     & $^{141}$Ba    & $^{142}$Ba     & $^{143}$Ba      &$^{144}$Ba&$^{145}$Ba& $^{146}$Ba& $^{147}$Ba     & $^{148}$Ba    \\ \hline
\rule{0pt}{2.5ex}$^{100}$Mo     &                  &                &                &               & 0.004(1)$^{10}$&                 &          &          & 0.009(2) &                 &          \\
$^{101}$Mo                      &                  &                &                &               & &                &          &          & 0.017(3) &                 & 0.005(1)         \\
$^{102}$Mo                      &                  &                &                & 0.005(2)$^{9}$& 0.015(3)$^{8}$ & 0.023(5)        & 0.09(2)  & 0.12(2)  & 0.18(3)  & 0.18(3)         & 0.031(6) \\
$^{103}$Mo                      & 0.009(3)$^{11}$  & 0.004(1)$^{10}$ &\textless{}0.005(1)$^{9}$& 0.010(2)$^{8}$& 0.07(1)        & 0.20(3)         & 0.42(7)  & 0.44(7)  & 0.35(6)  & 0.21(4)         & 0.014(3) \\
$^{104}$Mo                      & 0.009(2)$^{10}$  & 0.007(2)$^{9}$ & 0.010(2)$^{8}$ & 0.05(1)       & 0.24(4)        & 0.53(8)         & 1.08(17) & 0.73(11) & 0.37(6)  & 0.18(3)         & 0.006(1) \\
$^{105}$Mo                      & 0.008(2)$^{9}$   & 0.003(1)$^{8}$ & 0.022(4)       & 0.18(3)       & 0.48(8)        & 1.01(17)        & 1.09(19) & 0.47(9)  & 0.10(2)  & \textless{}0.04 &          \\
$^{106}$Mo                      & 0.015(3)$^{8}$   & 0.010(2)       & 0.09(2)        & 0.39(7)       & 1.05(17)       & 1.09(17)        & 0.72(12) & 0.27(4)  & 0.02(1)  &                 &          \\
$^{107}$Mo                      & 0.009(2)         & 0.020(4)       & 0.13(3)        & 0.31(6)       & 0.57(10)       & 0.30(6)         & 0.16(3)  &          &          &                 &          \\
$^{108}$Mo                      & 0.011(3)         & 0.028(5)       & 0.14(3)        & 0.22(3)       & 0.22(4)        & \textless{}0.09 & 0.05(1)  &          &          &                 &          \\
$^{109}$Mo                      & 0.008(2)         & 0.015(3)       & 0.034(7)       &               &                &                 &          &          &          &                 &          \\
$^{110}$Mo                      & \textless{}0.008 & 0.008(2)       &                &               &                &                 &          &          &          &                 &         
\end{tabular}
 \end{ruledtabular}
\end{table*}

\begin{figure}[!hbp]
    \centering
    \includegraphics[width=\columnwidth]{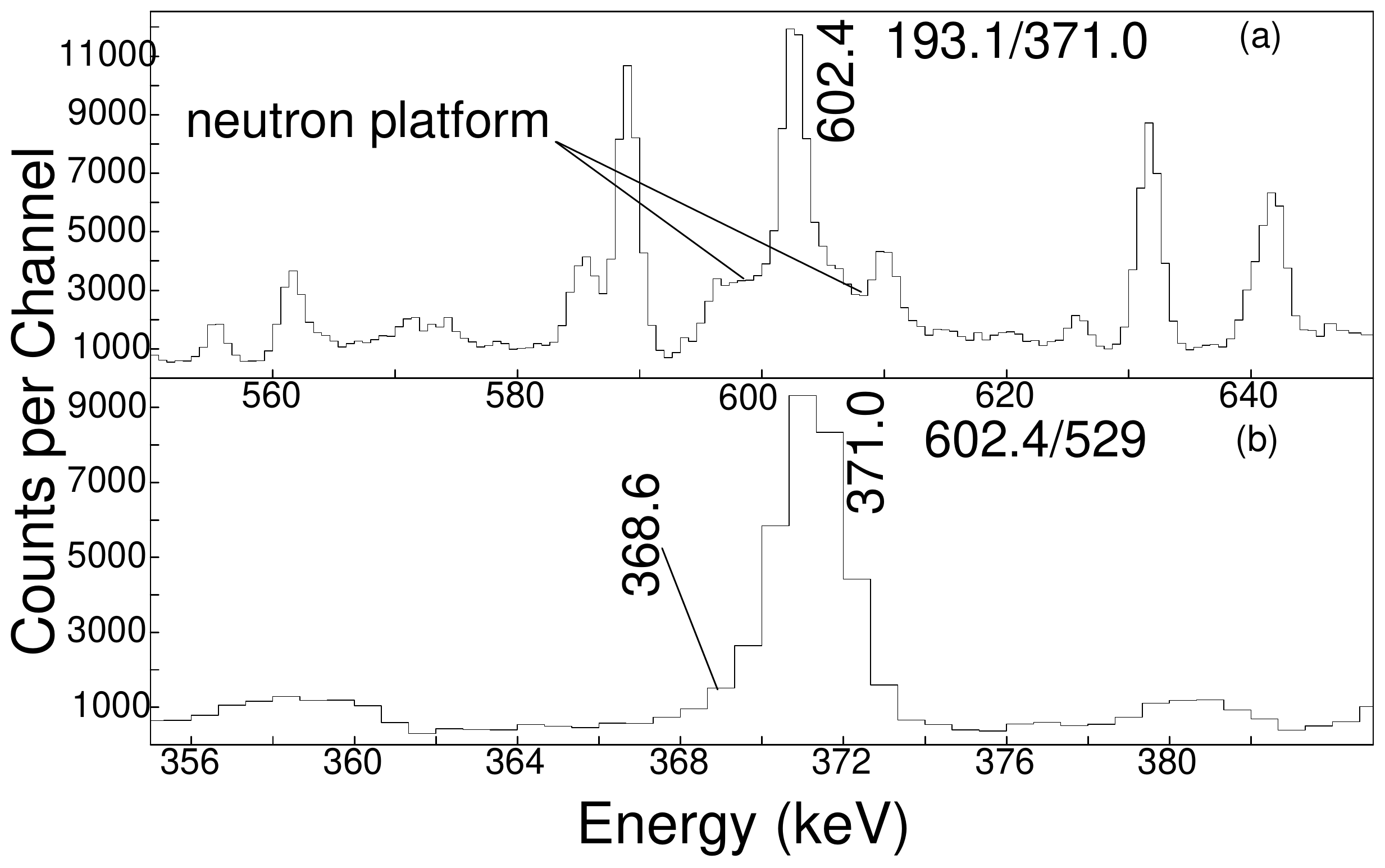}
    \caption{\label{yield-spectrum} Gamma-ray coincidence spectra by gating on (a) 193.1 and 371.0 keV transitions in $^{108}$Mo to show the neutron inelastic scattering platform, and (b) 602.4 and 529 keV transitions to show the 369 keV in $^{104}$Mo and 371 keV in $^{108}$Mo. Part (b) shows difference between Fig. 1 in Ref. \cite{Bis} using the same gate. See text for more details. }
    \label{fig:my_label1}
\end{figure}

The new yields of Ba-Mo are given in Table~\ref{mp-ba} and Fig.~\ref{yield-curves}. The 8-10 neutron yields presented in the present study are much lower than both the ones reported earlier; contributing $\sim$1.5(4)$\%$ of the first mode. In the first report~\cite{Ter}, the second mode was reported to contribute $\sim$7$\%$ of the first mode with significantly lower TKE, 153/189 MeV \cite{Ter}. The second report to have observed this mode~\cite{Wu}, reported that it contributed $\sim$3$\%$. The current experimental data have improved statistics over the other two experimental data from which the first and second analysis came. Therefore, one would expect that the second mode would be more pronounced in this experiment. However, this is not the case because with improved statistics comes more complete level schemes that provide new insights on possible contamination that were otherwise not considered in the previous analyses causing either overestimation or underestimation of the yields. Gating on Ba isotopes and Mo isotopes should give the similar yield results. Such cross-checks were used in this experiment to investigate the contamination given that contaminates are more common in Ba-Mo than in Ce-Zr, Te-Pd and Nd-Sr pairs.

\begin{figure}[ht!]
    \centering
    \includegraphics[width=\columnwidth]{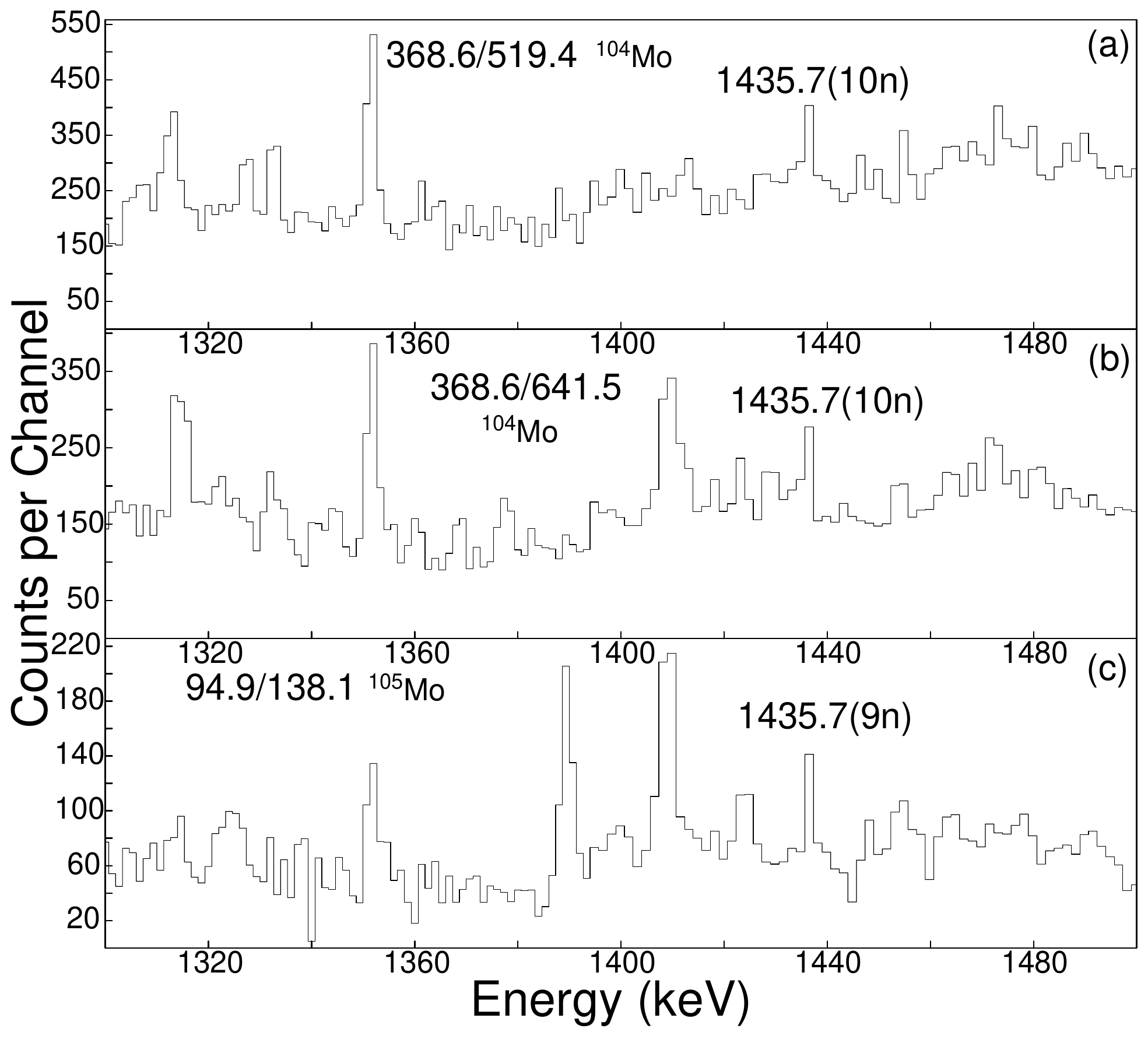}
    \caption{\label{yield_evi10}Gamma-ray coincidence spectra by gating on (a) 368.6 and 519.4 keV transitions in $^{104}$Mo to show evidence for the 10 neutron channel at 1435.7 keV in $^{138}$Ba and in (b) another gate on 368.6 and 641.6 keV transitions in $^{104}$Mo to give further evidence of the 10 neutron channel in the $^{138}$Ba-$^{104}$Mo. In (c) a gate on 94.9 and 138.1 keV transitions in $^{105}$Mo to show evidence for the 9 neutron channel at 1435.7 keV in $^{138}$Ba for the Ba-Mo pair.}
    \label{fig:my_label}
\end{figure}

In detail, in the analysis of Mo-Ba yields, one has to be extra careful when determining the yields of $^{140}$Ba-$^{104}$Mo and the $^{138}$Ba-$^{104}$Mo which correspond to the rare 8 and 10 neutron channels and the $^{140}$Ba-$^{108}$Mo and the $^{138}$Ba-$^{108}$Mo yield which correspond to the 4 and 6 neutron channels. This is because of the possible contamination that arise from the unresolved 192.4 keV and 192.9 keV 2$^{+}$ $\rightarrow$ 0$^{+}$ transition for $^{104}$Mo and $^{108}$Mo, respectively (see \cite{Bis} for similar analysis). In a previous analysis \cite{Bis}, a gate on 602.4/529 keV in $^{140}$Ba was used to measure the intensities of 368.6 keV and 371.0 keV transitions in $^{104}$Mo and $^{108}$Mo, respectively. In such gate, the 369 keV peak has $\sim$ 30 counts and is 1/3 (1/14 in our data) of the 371 keV one. In contrast, as seen in Fig. \ref{yield-spectrum} part (b), our data show $\sim$ 9000 counts for the 371 keV peak, while the 369 keV one is just above the background. Thus, it is possible that the $^{140}$Ba-$^{104}$Mo yield was overestimated due to the background fluctuation (10-20 counts) in Ref. \cite{Bis}. A gate was set on the first two transitions of the $^{108}$Mo isotope and the ground state transition (1435.7 keV) of $^{138}$Ba was measured as well the (8$^{+}$ $\rightarrow$ 6$^{+}$) transition populating the isomeric state at 2089 keV level because it is very strong in our data. When measuring the yields of $^{138}$Ba-$^{104}$Mo, however, the 192.4 keV from $^{104}$Mo has to be avoided to prevent contamination from 192.9 keV from $^{108}$Mo since it is strong and can enhance this yield. Instead, gates on 368 keV and 519 keV transitions from $^{104}$Mo were set and this time only the 1435.7 keV transition in $^{138}$Ba was measured (see Fig.~\ref{yield_evi10} (a)). Also in Fig.~\ref{yield_evi10} (b) is shown the 9 neutron channel seen in the 94.9-138.1 keV transitions in $^{105}$Mo to show the 9 neutron channel at 1435.7 keV in $^{138}$Ba.

Furthermore, unlike in Ref. \cite{Bis} we did not set a gate using the 602 keV from $^{140}$Ba when determining the yield of $^{140}$Ba-$^{108}$Mo pair because 602 keV is present in $^{104}$Mo from the 8$^{+}$ $\rightarrow$ 6$^{+}$ transition feeding into the 1725 keV level and another weaker 602 keV 8$^{+}$ $\rightarrow$ 6$^{+}$ transition feeding into the 2685.4 keV level. This means that setting any gate with 602 keV from $^{140}$Ba to measure desired peaks in $^{104}$Mo would bring in contamination. The other reason is that 602 keV lies on a complex region associated with an inelastic neutron scattering in germanium of the detectors as discussed in Ref.~\cite{Swu}. This neutron platform is not negligible; the background around this region is too high and as a result it is in coincidence with every other peak on the spectra (see Fig.~\ref{yield-spectrum} (a)). In Ref.~\cite{Swu}, the 528.2 keV 4$^{+}$ $\rightarrow$ 2$^{+}$ transition was used in the place of the 602.4 keV in $^{140}$Ba. However, transitions with energies close to 528 keV are present in $^{104}$Mo from 7$^{-}$ $\rightarrow$ 6$^{+}$ (feeding into the 2083.8 keV level), $^{105}$Mo from 21/2$^{-}$ $\rightarrow$ 19/2$^{-}$ (feeding into the 1352.9 keV level), $^{106}$Mo from 6$^{+}$ $\rightarrow$ 6$^{+}$ (feeding into the 1033.48 keV level), and $^{108}$Mo from 6$^{+}$ $\rightarrow$ 4$^{+}$ (feeding into the 564 keV level). Although they are weak transitions, when considering which one to gate on between the $^{104}$Mo and $^{140}$Ba, they are comparable in intensities when gating on $^{104}$Mo which would result in contamination but sufficient when gating on $^{140}$Ba. Such cases should also be carefully treated when measuring other high neutron channels with low yields, e.g $^{105,106}$Mo-$^{140}$Ba pairs.

Another approach that has been used in the past to resolve this problem is presented in Ref. \cite{Goo}. In the analysis of Ref. \cite{Goo}, the intensities of 519 keV (6$^{+}$ $\rightarrow$ 4$^{+}$ transition in $^{104}$Mo) and the 414 keV (4$^{+}$ $\rightarrow$ 4$^{+}$ transition in $^{108}$Mo) were measured instead of the ground state transitions for the yields. However, there is another 414 keV present and strong in $^{107}$Mo from (15/2)$^{-}$ $\rightarrow$ (11/2)$^{+}$ (feeding into the 2083.8 keV level). When gating on Ba transitions to measure the 414 keV in $^{108}$Mo, the strong 414 keV transition from $^{107}$Mo will contaminate the spectra. Whereas, the 519 keV transition is okay in this case because even though it is present in $^{103}$Mo it is weaker to contaminate the spectrum. Therefore, a gate on (519/641) keV from $^{104}$Mo was used to measure the 602 keV in $^{140}$Ba. These two gated transitions are located high enough in the $^{104}$Mo level scheme and have no feeding from the two contaminants (602 keV transitions mentioned earlier) in $^{104}$Mo. In this case, we used a local background subtraction which was set higher than usual to reduce the contribution from the neutron platform. This too does not completely circumvent the problems but it gave us a good approximation of what the yield should be. For more major overlapping transitions in Ba-Mo pairs to be considered when conducting this analysis refer to Table~\ref{trans}. Through this thorough examination of the Ba-Mo yield there is clear evidence of the 9 and 10 neutron channel yields as in Fig.~\ref{yield_evi10}.
 
The errors are significantly reduced because of the improved statistics, the use of quadruple coincidence data and improved knowledge of level schemes. To calculate all absolute errors, the experimental data were normalized to values from Wahl's tables \cite{Wahl}. Specifically, the summation of $^{144}$Ba yields was normalized to Wahl's value because it was the strongest yield in our experiment. Note that the values from Wahl's tables only considered ground state $\gamma$ transitions but we have considered the branching ratios from feeding bands. The 15$\%$ errors from Wahl's data were added to our absolute errors as well as 5-10 $\%$ experiment errors from missing transitions and contamination in our data.

   \begin{figure}[ht!]
    \centering
    \includegraphics[width=\columnwidth]{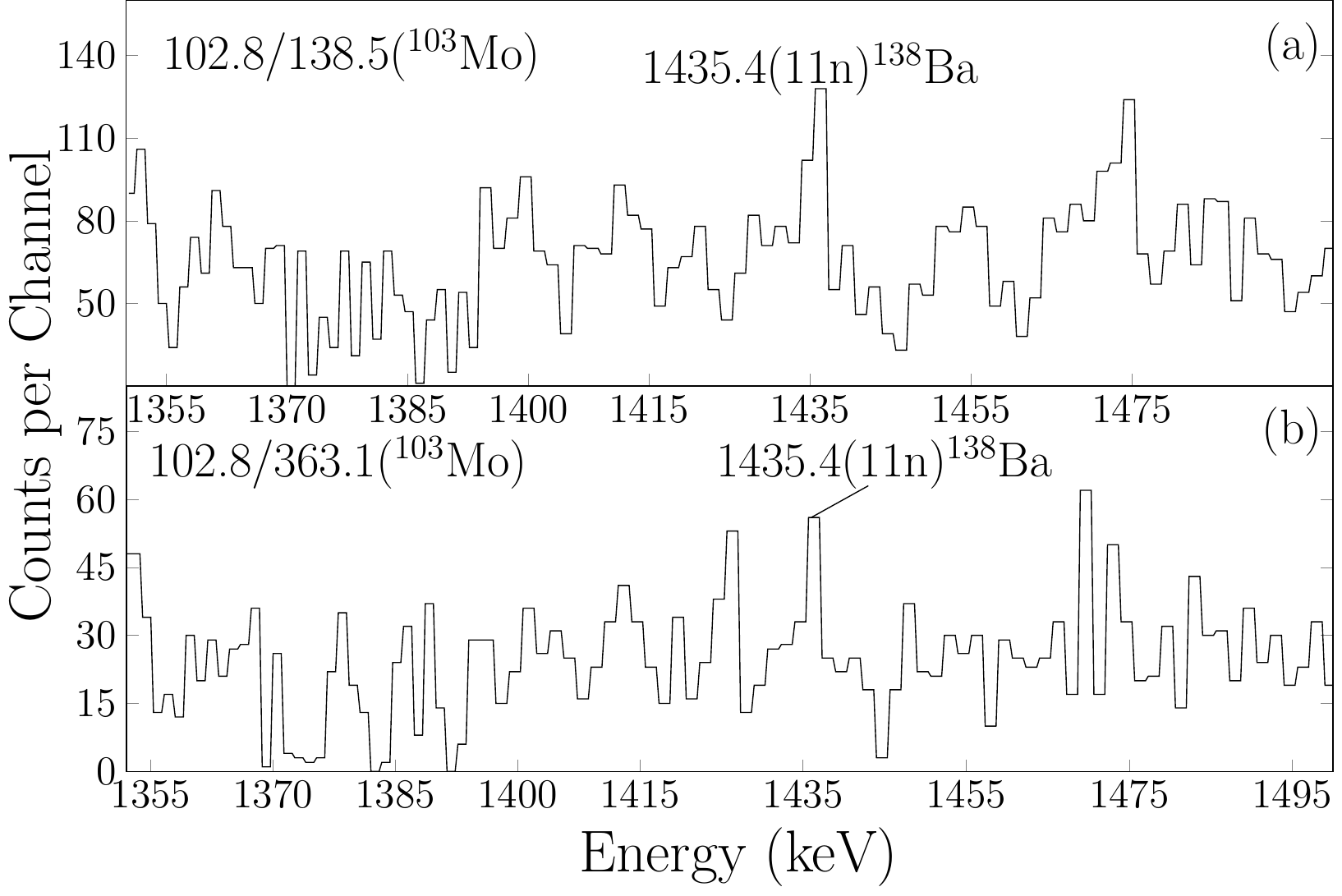}
    \caption{\label{11neutron} Gamma-ray coincidence spectra by gating on (a) 102.8 and 135.5 keV transitions in $^{103}$Mo to show evidence for the 11 neutron channel at 1435.7 keV in $^{138}$Ba and in (b) another gate on 102.8 and 363.1 keV transitions in $^{103}$Mo to give further evidence of the 11 neutron channel in the $^{138}$Ba-$^{104}$Mo pair.}
    \label{fig:my_label1}
\end{figure}

\begin{table}[!htp]
\caption{\label{trans} Part of the major overlapping energies transitions in Ba-Mo pairs that could result in contamination. See text for more instructions.}
\resizebox{\columnwidth}{!}{
\begin{tabular}{|c|l|l|}
\hline
Energy (keV) & Nuclei           & $E_i$ to $E_f$ (keV) \\ \hline
\rule{0pt}{2.5ex}110          & $^{107}$Mo            & 458$\rightarrow$348              \\
             & $^{109}$Mo            & 333$\rightarrow$222              \\
             & $^{145}$Ba            & 618$\rightarrow$508              \\
             & $^{147}$Ba            & 110$\rightarrow$0                \\ \hline
\rule{0pt}{2.5ex}113          & $^{103}$Mo            & 354$\rightarrow$241              \\
             & $^{145}$Ba            & 113$\rightarrow$0                \\ \hline
\rule{0pt}{2.5ex}172          & $^{105}$Mo            & 796$\rightarrow$623              \\
             & $^{106}$Mo            & 172$\rightarrow$0                \\
             & $^{107}$Mo            & 492$\rightarrow$320              \\ \hline
\rule{0pt}{2.5ex}185          & $^{145}$Ba            & 463$\rightarrow$277              \\
             & $^{147}$Ba            & 185$\rightarrow$0                \\ \hline
\rule{0pt}{2.5ex}192          & $^{104}$Mo            & 192$\rightarrow$0                \\
             & $^{108}$Mo            & 193$\rightarrow$0                \\
             & $^{138}$Ba            & 2089$\rightarrow$1898            \\ \hline
\rule{0pt}{2.5ex}250          & $^{103}$Mo            & 354$\rightarrow$103              \\
             & $^{147}$Ba            & 360$\rightarrow$110              \\ \hline
\rule{0pt}{2.5ex}414          & $^{107}$Mo            & 566$\rightarrow$152              \\
             & $^{108}$Mo            & 979$\rightarrow$564              \\ \hline
\rule{0pt}{2.5ex}493          & $^{107}$Mo            & 950$\rightarrow$458              \\ 
             & $^{143}$Ba           & 954$\rightarrow$461              \\ \hline
\rule{0pt}{2.5ex}519          & $^{103}$Mo            & 1157$\rightarrow$637             \\
             &             & 1180$\rightarrow$561             \\ \hline
\rule{0pt}{2.5ex}529          & $^{104}$Mo            & 2612$\rightarrow$2083            \\
             & $^{105}$Mo            & 1882$\rightarrow$1353            \\
             & $^{106}$Mo            & 1563$\rightarrow$1033            \\
             & $^{108}$Mo            & 1508$\rightarrow$979             \\
             & $^{140}$Ba            & 1130$\rightarrow$602             \\
             &             & 1660$\rightarrow$1130            \\ \hline
\rule{0pt}{2.5ex}602          & $^{104}$Mo            & 2326$\rightarrow$1725            \\
             &                  & 2685$\rightarrow$2083            \\
             & $^{140}$Ba            & 602$\rightarrow$0                \\
             & neutron platform &      \\ \hline                 
\end{tabular}}
\end{table}
As seen in Fig.~\ref{yield-curves}, a similar deviation from a Gaussian fit to the data for the 0 to 7 neutron emission channels is seen at neutron numbers 7, 8, 9 and 10 in the Ba-Mo yields as observed in \cite{Ter, Wu}. In comparison to these results, a noticeable difference is that in the present analysis we have a more complete set of yield pairs; $^{100-110}$Mo and $^{138-148}$Ba. This is not the case for the earlier analyses where $^{139}$Ba is missing in \cite{Ter, Bis, Wu} and $^{138}$Ba in \cite{Bis}. These are very important components of the analysis as they contribute to the intensity of the second hot mode. Additionally, the 9 and 10 neutron channels were not reported in \cite{Bis} and \cite{Goo, Ram} (same data set in these two) did not report only the 10 neutron channel. Note that there is a typographical error in the $^{142}$Ba-$^{102}$Mo yield in Ref. \cite{Goo}. This reported yield is too small (0.007) compared to $^{144}$Ba-$^{104}$Mo (102) in the same reference. The second smallest reported yield in Ref.~\cite{Goo} was 0.35, which is two orders larger than the 0.007 value. However, as shown in Fig~\ref{yield_evi10}, the 9 and 10 neutron channels are present. And in the current study we have also observed the 11 neutron channel at the 1435 keV peak in $^{138}$Ba as seen in Fig.~\ref{11neutron}. This channel is observed in several gates but in Fig.~\ref{11neutron} we only show two gates on 102.8/138.5 keV peaks in (a) and 102.8/363.1 keV peaks in (b) and they are both from $^{103}$Mo. Also in Fig. \ref{tailscurves}, we show a second Gaussian fit to the 8, 9, and 10 as reported earlier \cite{And} and added the 11 neutron channel. In Ref.~\cite{And}, we fitted a second Gaussian by means of restricting the peak position of the second mode to greater than 6 neutrons emitted. However, in the present study we were able to obtain a reasonable fit by restricting the peak position of the second fit to $\sim 8$ and width of the second curve was fixed to the width of the first curve. This new analysis of Ba-Mo fission pairs, coupled with the new analysis of Ce-Zr yields, which shows a reduced ``extra hot mode'', and the Te-Pd and Nd-Sr yields, which do not exhibit 8, 9, 10 neutron emissions, confirms the existence of this ``extra hot mode" in the Ba-Mo and now found in Ce-Zr yields.

 \section{Conclusion}
 
 In the present work, new yield matrices were determined for Te-Pd, Nd-Sr, Ce-Zr and Ba-Mo fission partners from the spontaneous fission of $^{252}$Cf. A similar deviation from the Gaussian fit to the normal fission mode was found in Ba-Mo for the 8, 9, and 10 neutron channels as found in previous analyses to confirm the existence of the proposed ``extra-hot-fission'' mode. We have also observed an ``extra hot fission mode'' for the first time in Ce-Zr pairs. The observation of these modes in both pairs can be explained by considering that $^{143,144,145}$Ba and $^{146,148}$Ce have been determined to be octupole deformed which can help give these nuclei high internal energy at scission and in turn gives rise to high neutron multiplicities. This is in addition to the possible hyperdeformation suggested for these nuclei. Errors are reduced in this newest analysis compared to previous studies because of the greater statistics of the latest Gammasphere experiment and the use of quadruple coincidences in the analysis and improved level schemes. A new experiment is being planned to do fission fragment-$\gamma$-$\gamma$ coincidence studies to investigate details of the fission process and to study new more neutron-rich nuclei. In addition, the investigation will study the existence of an ``second extra hot mode'' observed in Ba-Mo  and Ce-Zr fission yields as well as ascertain whether these second modes are a result of hyperdeformation and/or octuple deformation of $^{144,145,146}$Ba and $^{146,148}$Ce.
 
\begin{acknowledgments}
The work at Vanderbilt University and Lawrence Berkeley National Laboratory are supported by the US Department of Energy under Grant No. DE-FG05-88ER40407 and Contract No. DE-AC03-76SF00098. The work at Tsinghua University was supported by the National Natural Science Foundation of China under Grant No. 11175095. The work at JINR was partly supported by the Russian Foundation for Basic Research Grant No. 08-02-00089 and by the INTAS Grant No. 2003-51-4496.
\end{acknowledgments}

\end{document}